\documentclass[12pt,preprint]{aastex}

\shorttitle{HE~1015$-$2050: A HdC STAR AT HIGH GALACTIC LATITUDE}
\shortauthors{Aruna Goswami et al.}
\begin{document}
\title{HE~1015$-$2050: DISCOVERY OF A HYDROGEN-DEFICIENT CARBON STAR AT HIGH GALACTIC LATITUDE}
\author{ARUNA GOSWAMI\altaffilmark{ }, DRISYA KARINKUZHI\altaffilmark{ } }
\affil{Indian Institute of Astrophysics, Bangalore 560034, India\\
 e-mail: aruna@iiap.res.in
}

\and

\author{N. S. SHANTIKUMAR\altaffilmark{ }}
\affil{Centre for Research and Education in Science and Technology, Indian Institute of Astrophysics, Shidlaghatta Road, Hosakote 562114, India\\}

\begin{abstract}
Medium resolution spectral analysis of  candidate Faint High Latitude 
Carbon (FHLC) stars from Hamburg/ESO survey has given us the potential 
to discover objects of  rare types.  Two primary spectral characteristics
 of R  Coronae Borealis (RCB) stars are hydrogen deficiency and weaker 
CN bands relative to C$_{2}$ bands. They are  also 
characterized by their characteristic  location in the J-H, H-K plane 
with respect to cool carbon stars. From a  spectral analysis of a sample 
of  243 candidate FHLC stars, we have discovered a hydrogen-deficient 
carbon  (HdC) star HE~1015$-$2050,  at high Galactic latitude. A differential 
analysis of its  spectrum  with that of the spectrum of
U Aquarii (U~Aqr), a well-known cool HdC  star of RCB type,
provides sufficient evidence to put this object  in a group  same as that
of  U~Aqr. Further, it is  shown that HE~1015$-$2050 does not belong 
to any of the C-star  groups CH,  C-R, C-N or  C-J. Cool RCB  stars form 
a group of relatively rare astrophysical objects; approximately 51 are 
known in the Galaxy and some 18 in the Large Magellanic Clouds (LMC) and five
 in Small Magellanic Cloud (SMC). The present discovery adds  a new  member 
to this rare group. Although its spectral characteristics and its 
location in the J-H vs H-K plane  places HE~1015$-$2050 in the same group
to which  U~Aqr belongs, extended photometric observations would be 
useful to learn  if there is any sudden decline in brightness, this 
being a characteristic  property of HdC stars of RCB type. 

\end{abstract}


\section{INTRODUCTION}

HdC stars are spectroscopically similar to 
the R Coronae Borealis RCB type stars. While they may show  small-amplitude light variations 
\citep{Lawson97}  the deep minima and infrared excesses that 
are characterististics of RCB stars are absent in these stars 
\citep{Warner67,Feast73,Feast97}. These
two groups have been collectively identified as HdC stars \citep{Warner67}. 
Inspite of several studies, the formation mechanism(s) of these stars still 
remains  poorly understood. Insight into this  can be obtained
 through detailed studies of a  large  statistically significant 
 number of  these objects.  However,  as these  objects are  rare,
in particular, only about 51 Galactic RCB stars are known so far;
 a search  for these objects  seems therefore extremely worthwhile.

HD~182040 is the first object to be identified as HdC star \citep{Bidelman53}.  
The recent surveys of the LMC \citep{Alcock01,Morgan03} 
revealed that,  unlike in the
Galaxy, RCB star population is  dominated by cool stars  rather than 
warm stars by 7 : 2.  As the environment of LMC is more metal-poor 
than the solar neighbourhood,  it seems worthwhile to search for HdC 
stars in  the environment of lower metalicity (sub solar). The sample
of FHLC stars from Hamburg survey \citep{Christlieb01} offers such 
a possibility being objects with small initial mass and possible
lower metallicity. Medium resolution spectral analyses of   about 243
objects from this sample of 403 show that  161 of them exhibit
strong C$_{2}$ bands in their spectra; their spectral characteristics
are discussed in several papers \citep{Goswami05,Goswami06,Goswami07,Goswami10a,Goswami10b}. In this paper we discuss the spectrum of
HE~1015-2050; a careful inspection  reveals the star to exhibit 
spectral characteristics of cool HdC stars. Further, a differential  
analysis of the star's spectrum with the spectrum of U~Aqr, provides  
sufficient evidence to put this  object  in a group same as U~Aqr. 
Apart  from the very large strengths  of Sr II at 4077 \AA\,  and 
Y II at 3950 \AA\,, U~Aqr is a typical RCB  type variable and its high 
radial velocity (103 $\pm$ 20 km s$^{-1}$)  indicates that it 
belongs to an old population and that its  main sequence progenitor 
had a mass $\sim$ 1 M$_{\odot}$ \citep{Bond79}. These two features of 
Sr II and Y II  are also observed in the spectrum  of  HE~1015-2050 
and are very   similar to those in U~Aqr.

\section{OBSERVATIONS AND DATA REDUCTION}

The star  HE~1015$-$2050 was   observed with 2-m Himalayan Chandra Telescope
(HCT)  at the Indian Astronomical Observatory (IAO), Mt. Saraswati, 
Digpa-ratsa Ri, Hanle   on  2009 December 5 and 2010 january 24, 
using Himalayan Faint Object Spectrograph Camera (HFOSC). The grism used 
in the present investigation and camera combination yielded a resolution 
of $\sim$ 1330  (${\lambda}/{\delta\lambda}$) covering a spectral interval 
of  3800 to 6800 \AA\,. Spectra of HD~209621, HD~182040, U~Aqr, V460~Cyg 
and ES~Aql used for comparison  were obtained during earlier observation 
cycles using the same observational set up. A spectrum of C-R star 
HD~156074 from \citet{Barnbaum96} atlas, which has a comparable 
resolution to the present observations have been used for comparisons. 
Wavelength calibration was accomplished using observations of  Th-Ar  
hollow cathod lamp.  The  CCD data were reduced  using the IRAF software 
spectroscopic reduction  packages. The photometric parameters of
HE 1015-2050 and the comparison HdC and RCB stars are listed in Table 1.

\section{ SPECTROSCOPIC CONFIRMATION}

The membership of a star to a particular group can be determined in terms 
of a number of parameters, spectral characteristics being an important one.
In terms of spectral characteristics  the spectra of cool RCB stars  
show distinct  C$_{2}$ molecular bands; these bands are  weakly visible 
in the warm  RCB stars. In Figures 1-3  we show a comparison  of 
the  spectrum of  HE~1015$-$2050  with the spectra of  two cool RCB type 
stars U~Aqr and ES~Aql, a well-known  non-variable HdC star HD~182040, 
a well-known CH star HD~209621 and  a C-N star V460~Cyg  in  the  
wavelength regions 3850-4950, 4900-6000 and  5950-6700 \AA\,, respectively.
Some of the key spectral features considered for classification of the 
program stars are the strength of G-band of CH around 4310 \AA\,
presence or absence of  isotopic bands of C$_{2}$ and CN, the Swan bands
$^{12}$C$^{13}$C and $^{13}$C$^{13}$C near 4700 \AA\,,  other C$_{2}$
bands in the 6000-6200 \AA\, region and the $^{13}$CN band near  6350 \AA\,.
In particular we have used  hydrogen-deficiency and the relative strength 
of C$_{2}$ bands in the 6000-6200 \AA\, region and the CN  bands near  
6206 \AA\, and 6350 \AA\, as  important classification criteria. As the
Balmer lines are weak in carbon stars, the strength/weakness of CH band 
in C-rich stars provides a measure of the degree of hydrogen deficiency. 

\subsection {Spectral characteristics: Hydrogen Deficiency}

An inspection  of the  spectra in the   G band  region (Figure 1) shows 
that this feature which is strongly seen in CH and  C-R stars' spectra 
is marginally detected in the spectrum of HE~1015$-$2050. The G band 
strength is about equal or weaker relative to its counterpart in the 
spectra of the HdC star HD~182040 and the two HdC star of RCB type 
U~Aqr and ES~Aql. H$_{\beta}$ feature,  which is generally seen in the 
spectra of  C-R and C-N stars, is not clearly detectable in the spectra  
of the   HdC star HD~182040, U~Aqr, ES~Aql  and  HE~1015$-$2050. Marginal 
detection or  non-detectability of  H$_{\beta}$ feature  could also be  
one of the evidence for hydrogen deficiency in U~Aqr, ES~Aql, HD~182040 
and  HE~1015$-$2050. 

H$_{\alpha}$ feature  which is distinctly seen in the spectra of CH star
HD~209621 and C-R star 156074 is also  not detectable  in HE~1015$-$2050,
HD~182040, U~Aqr and ES~Aql (Figure 3). In the case of some cool carbon stars, 
if not hydrogen poor, the Balmer lines show up in chromospheric emissions.  
Such chromospheric feature is  absent  in the spectrum of HE~1015$-$2050. 
This again hints at the hydrogen-poor nature of the  star.

\subsection{Spectral characteristics: Molecular bands of C$_{2}$ and CN}

One of the spectral  characteristics which seems to be shared by all cool 
HdC stars of  RCB type including U~Aqr is that the red CN bands are usually  
much weaker even in stars with strong C$_{2}$ bands \citep{Lloyd91,Alcock01,Morgan03}. These characterstics 
differentiate RCBs  from other carbon stars. The spectrum of HE~1015$-$2050 
is dominated by the blue degraded Swan system  of C$_{2}$ bands. 
As in the case of U~Aqr, carbon molecular band head around 6191 \AA\, is 
seen strongly in HE~1015$-$2050; this feature is much weaker  in HD~182040. 
As seen in  Figure 3, in the spectrum of C-N star  V460~Cyg the
C$_{2}$ bands around 6000-6200 \AA\, and CN bands around 6206 \AA\, 
and 6350 \AA\, appear almost of equal strength.   Although C$_{2}$ band 
depths  in the spectra of HE~1015$-$2050 and the two HdC stars of RCB type
are similar to  those in V460~Cyg, the red degraded CN bands appear much 
weaker. CN bands are clearly weaker in U~Aqr's spectrum; in the spectrum of
HE~1015$-$2050 these  bands are even weaker than  in U~Aqr. The presence of 
(6,2) ${\lambda}$6500 CN band is unlikely. Metal lines near 6500 \AA\, 
including Ba II at 6496 \AA\, and also a C$_{2}$ band (4,7) ${\lambda}$6533, 
are likely to contribute to the appearance of the spectrum around that 
region. The bands (4,0) CN ${\lambda}$6206 (Figure 3),
(6,1) CN ${\lambda}$5730 and (7,1) CN ${\lambda}$5239 (Figure 2) are  
almost absent in the HE star's spectrum; these bands are also very weak 
or absent in the normal RCB stars. The red isotopic bands involving 
$^{13}$C are not normally present in RCB stars \citep{Lloyd91}. 
As expected in RCB stars, the (1,0) $^{13}$C$^{12}$C $\lambda$4744 band 
is  present in HE~1015$-$2050, but  much weaker than the 
(1,0) $^{12}$C$^{12}$C $\lambda$4737 band. The RCB stars are reported 
to show the strong blue-degraded (0,1) CN 4216 member of the violet 
band system. The spectral region of HE~1015$-$2050 shortward of 4216 \AA\, 
bear a striking similarity with the spectrum of U~Aqr. The above spectral 
characteristics  can also be seen in other  cool Galactic  RCB  stars, e.g. 
Z UMi and S Aps \citep{Benson94,Goswami97a,Goswami97b,Goswami99}
and stand in favour of HE~1015$-$2050's  membership  to the cool RCB  
group. \\

\subsection{ Non-membership of  HE~1015$-$2050 in
CH, C-R, C-N and C-J group}

We have further examined the spectrum  of HE~1015$-$2050 for its possible
membership in any of CH, C-R, C-N, and C-J  groups. Following the 
carbon stars classification schemes and spectral criteria \citep{Keenan93,Barnbaum96,Keenan97,Goswami05}, we have 
shown that the star  HE~1015$-$2050 does not belong to any  of CH, C-R, 
C-N and  C-J groups. 

In the spectra of  CH and C-R stars, the CH band around 4310 \AA\,,
  H$_{\beta}$  and H$_{\alpha}$ features are  prominently seen. In 
Figures 1 and 3,  a  marginal or weak detection of these features in the 
spectrum of  HE~1015$-$2050 therefore is a strong indication  that it 
does not belong either to  CH or  C-R group. 

 The red degraded CN bands around (6,1) ${\lambda}$5730, 
(7,2) ${\lambda}$5860, (4,0) ${\lambda}$6206, (5,1) ${\lambda}$6335  are 
usually seen as prominently as the C$_{2}$ bands in C-N stars. Weaker 
CN bands  in the spectrum of  HE~1015$-$2050  suggest that the star is 
not a conventional  C-N star. As seen in Figure 1, Ba II at 4554 \AA\, 
and Sr I at 4607 \AA\, which are seen quite distinctly in the spectrum of 
the C-N star V460~Cyg  could only be marginally  detected in the spectra of
HE~1015$-$2050 and U~Aqr spectra. C-N stars are characterized by strong 
depression of light in the violet part of the spectrum (i.e., V460~Cyg in 
Figure 1). The cause of rapidly weakening continuum below about 4500 \AA\, 
is  believed to be due to scattering by particulate matter; but this idea 
is not yet fully established. Blue/voiolet part of the spectrum in 
HE~1015$-$2050 (and also of U~Aqr) is accessible to observation and 
atmospheric analysis. As seen from Figure 1, features due to Ca K, 
Ca H, Y II, \& Sr II that are not seen in the spectrum of V460~Cyg are 
observable with almost equal strengths in the spectra of U~Aqr and 
HE~1015$-$2050. In these respects  the star  HE~1015$-$2050 could not be 
classified as a C-N star.

The possibility of HE~1015$-$2050 belonging to C-J group is 
ruled out  by the absence of the  Blue-green Merrill-Sanford (MS) 
bands of SiC$_{2}$. MS red degraded bands are generally prominently seen 
in C-J stars. In the spectrum of HE~1015$-$2050, MS bands of SiC$_{2}$ at
4640 \AA\, and 4977 \AA\, both could not be detected. Thus we see that 
HE~1015$-$2050 does not belong to any of the C-star groups CH, C-R, C-N, 
and C-J .  

\subsection{ Atmosphere of HE~1015$-$2050}

The stellar atmosphere of HE~1015$-$2050  has comparable effective temperature
as those of cool RCB type stars. The estimated  effective temperature
(T$_{eff}$) of  HE~1015$-$2050 ${\sim}$ 5263 K derived using semiempirical 
temperature  calibration  relations offered by \citet{Alonso94,Alonso96, 
Alonso98} is similar to  the estimated $T_{eff}\,\,{\sim}$  5000 K for  cool
Galactic RCB stars known e.g., S Aps, WX CrA and U~Aqr \citep{Lawson90}. 

Carbon isotopic ratio $^{12}$C/$^{13}$C,  widely used as a mixing
diagnostics  provides an important  probe of stellar evolution.
These ratios measured on medium-resolution spectra, although not accurate,
provide  a fair indication of the evolutionary  stages. Using the molecular
band depths  of (1,0) $^{12}$C$^{12}$C ${\lambda}$4737 and
(1,0) $^{12}$C$^{13}$C ${\lambda}$4744 we have estimated
the ratio $^{12}$C/$^{13}$C  to be  ${\sim}$ 6.4 in HE~1015$-$2050
and ${\sim}$ 3.4 in U~Aqr.
In the spectrum of HD~182040, (1,0) $^{12}$C$^{13}$C${\lambda}$4744 band
is not observed.

{\it Atomic lines.}\,\,\,\, Similar to U~Aqr, the spectrum of HE~1015$-$2050 
is characterized  by an extraordinary strong line of Sr II  at 4077 \AA\,.
Y II line at 3950 \AA\, is distinctly prominent in the spectrum of 
HE~1015$-$2050. Although spectroscopically less prominent, this line
is  also considerably strengthened in U~Aqr. Although detected 
clearly, there is no significant enhancement of Ba II at 4554 and 
6496 \AA\,. The atomic line of Fe I at 4045 \AA\, is clearly detected in 
HE~1015$-$2050 as in U~Aqr; however Fe I line at 4063 \AA\, is not clearly 
detected (Figure 1). Ca I line at 4226 \AA\,, distinctly detected in 
HD~182040, is not prominently seen in the spectra of HE~1015$-$2050 and U~Aqr.
Unlike HE~1015$-$2050 and U~Aqr, the spectra of HD~182040 and ES~Aql do 
not show lines of  Y II at 3950 \AA\, and Sr II  at 4077 \AA\,. The 
features due to  Na I D,  although  appearing  strong  
in HE~1015$-$2050 and  U~Aqr, are  much weaker relative to  their 
counterparts in V460~Cyg and HD~182040. The  H$_{\alpha}$ feature is  
not detected. The most striking feature in the spectra of U~Aqr
and HE~1015$-$2050 is the strength of the Sr II ${\lambda}$ 4215 line;
this feature  is  inextricably blended with the nearby
strong blue-degraded (0,1) CN 4216 band head in HD~182040.

\subsection{Location of HE~1015$-$2050 in J-H vs H-K plot}

Apart from the spectroscopic criteria infrared colors  from JHK photometry
may be used as  a supplementary diagnostic for identifying peculiar or 
rare stars. The Two Micron All Sky Survey (2MASS) measurements  \citep{Skrutskie06}  place the 
candidate cool  HdC star HE~1015$-$2050 on  the J-H versus H-K plane, at 
the bottom of the figure along with the cool  LMC RCB stars and near U~Aqr
(Figure 4).
The position of HE~1015$-$2050 is marked with an open circle symbol. The  
boxes representing the locations of CH stars ( thick solid line box) 
and C-N stars (thin solid line box) are taken from \citet{Totten00}. 
In this  figure, the locations of the  cool LMC RCB stars are shown with 
solid circles, and the comparison stars with solid squares.
The LMC photometry is taken from \citet{Alcock01}. The location of 
HE~1015$-$2050 in the vicinity of  U~Aqr in the J-H, H-K plane supports 
its identification with the group same as that of U~Aqr.

\section { CONCLUDING REMARKS}
Spectral analysis of  243 candidate FHLC stars of  \citet{Christlieb01}, revealed one star HE~1015$-$2050, to be a bona-fied member of 
HdC stars. Its  spectrum  satisfies  the primary spectral characteristics 
of HdC stars of RCB type. In particular, its  spectrum  resembles 
closely the spectrum of  U~Aqr,  a cool Galactic HdC star of  RCB type. 

In the JHK diagram the star  occupies a position along with the cool  RCB 
stars.  This is consistent with the fact that the spectral characteristics 
of HE~1015$-$2050 do not match those of CH, C-R,  CN, or C-J stars.

HdC stars and in particular RCB type stars are  a rare class of objects; 
only about  51 RCB  stars are known in our Galaxy, about 18 in LMC and 
five in SMC  \citep{Feast72,Alcock01,Morgan03,Tisserand04,Tisserand08}.  
Discovery of HE~1015$-$2050 adds a new member to this  rare class of stars. 
\citet{Lawson90} suggested that the apparent lack of cool RCBs is a
selection effect, and that the true number of cool RCB stars (${\le}$ 5000 K)
may greatly exceed the number of warm RCB (${\sim}$ 7000 K) stars.
 The recent surveys of the LMC \citep{Alcock01,Morgan03} revealed that RCB star population is dominated 
by cool stars rather than warm stars by 7 : 2 unlike the Galaxy. The 
environment of LMC being more metal poor than the solar neighbourhood may
also be a contributing factor.  The program
stars  being high latitude objects,  with smaller initial mass and 
possible lower metallicity  offers potential to discover such rare objects.

The observed elemental abundances of  HdC and RCB stars are similar and 
seem to represent  the products of  a combination of H- and He-burning 
 \citep{Warner67,Asplund97,Asplund00,Kipper02}. 
 Current evolutionary models, however, 
cannot reproduce completely the typical RCB star abundances.
None of the RCB and HdC stars  is known to be binary \citep{Clayton96}.
 Abundances of  Rb, Sr, Y and Zr are 
greatly enhanced in U~Aqr, but  Ba does not show a  large enhancement. 
Estimates of \citet{Vanture99} ([Y/Fe] = +3.3, [Zr/Fe] = +3.0, and 
[Ba/Fe] = +2.1)
are larger than those estimated by \citet{Bond79}, but in general
agreement with those of \citet{Malaney85}.
 Such  abundance patterns resembles 
the weak component of the s-process found in solar system material
which is best described by a single neutron irradiation
\citep{Beer89}.
The similarities of the spectra of U~Aqr and HE~1015$-$2050 suggest that
the HE star is also a potential candidate for studying $s$-process
nucleosynthesis.

Low mass hydrogen-deficient stars are associated with late stage of
stellar evolution. RCB phenomena characterized by spectacular dimmings of
five or more magnitudes within  a few days  from the onset of minimum 
(followed by slow recovery to maximum light) is believed  to be 
due to directed mass-ejections and  primarily a signature of surface 
activity rather than chemical peculiarity. Absence of such irregular 
fadings may indicate absence of 
such mass-ejection  episodes; but, whether that is characteristic 
of a particular evolutionary stage is not yet established. Detailed  
spectroscopic  as well as  photometric studies  are expected to provide 
much insight  into these aspects. The spectral characteristics and its 
location in the J-H vs H-K plane certainly allow us to place 
HE~1015$-$2050 in the same group to which  U~Aqr belongs; however,
extended photometric observations  of this  object would be useful to 
learn the nature and extent of photometric variations. In  particular, 
it would be interesting to see if there is any sudden decline in 
brightness, this being a characteristic  property of HdC stars of 
RCB type.\\

D. K. is a  JRF in the DST project No.
SR/S2/HEP-09/2007;  funding from the project  is  greatfully acknowledged. 
This work made use of the SIMBAD astronomical database, operated at CDS, 
Strasbourg, France, and the NASA ADS, USA.\\

\begin{figure*}
\centering
\includegraphics[angle=0,height=13cm,width=15cm]{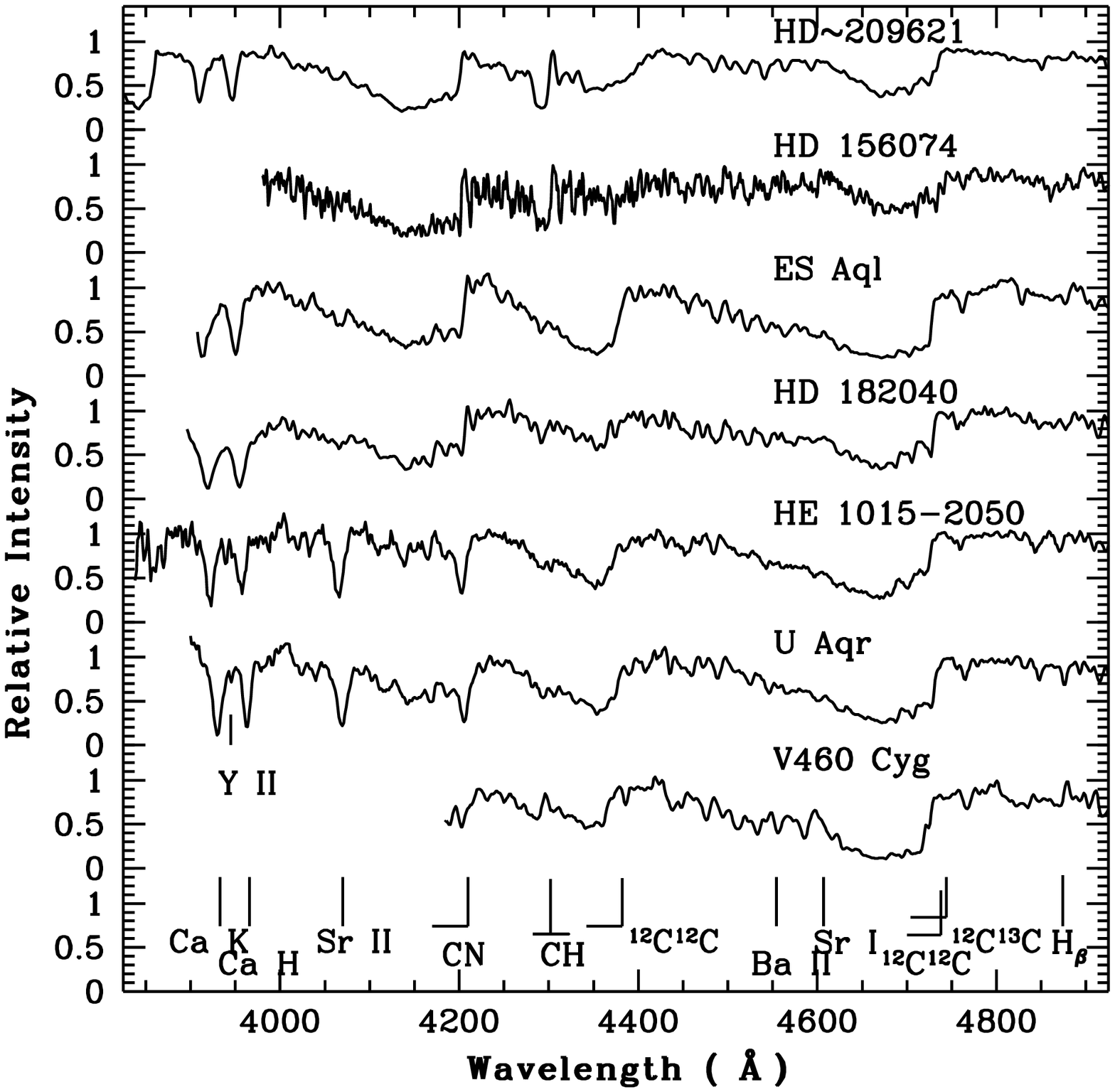}
\caption
 { A comparison between the spectrum of  HE~1015$-$2050 with the spectra 
of V460~Cyg (C-N star),  U~Aqr, ES~Aql (cool HdC stars of RCB type),  
HD~182040 (a non-variable HdC star), HD~156074 (C-R star), and  
HD~209621 (CH star)  in the wavelength region 3850-4950 \AA\,. 
G-band of CH distinctly seen in the CH and C-R stars' spectra are 
barely detectable in the spectra of HE~1015$-$2050 and other HdC stars 
spectra. The large enhancement of Sr~II at 4077 \AA\, in the spectrum 
of U~Aqr is easily seen to appear with almost equal strength in the 
spectrum of HE~1015$-$2050.  Y~II line at 3950 \AA\, is  detected in the 
spectra  of both HE~1015$-$2050 and U~Aqr. 
The most   striking  feature in the spectra of U~Aqr
and HE~1015$-$2050 is the strength of the Sr~II ${\lambda}$ 4215 line;
 this feature  is  inextricably blended with the  nearby  strong 
blue-degraded (0,1) CN 4216 band head in HD~182040. The spectrum of 
HE~1015$-$2050 compares closest to the  spectrum  of  U~Aqr.
\label {fig1}
} 
\end{figure*}

\begin{figure*}
\centering
\includegraphics[angle=0,height=13cm,width=15cm]{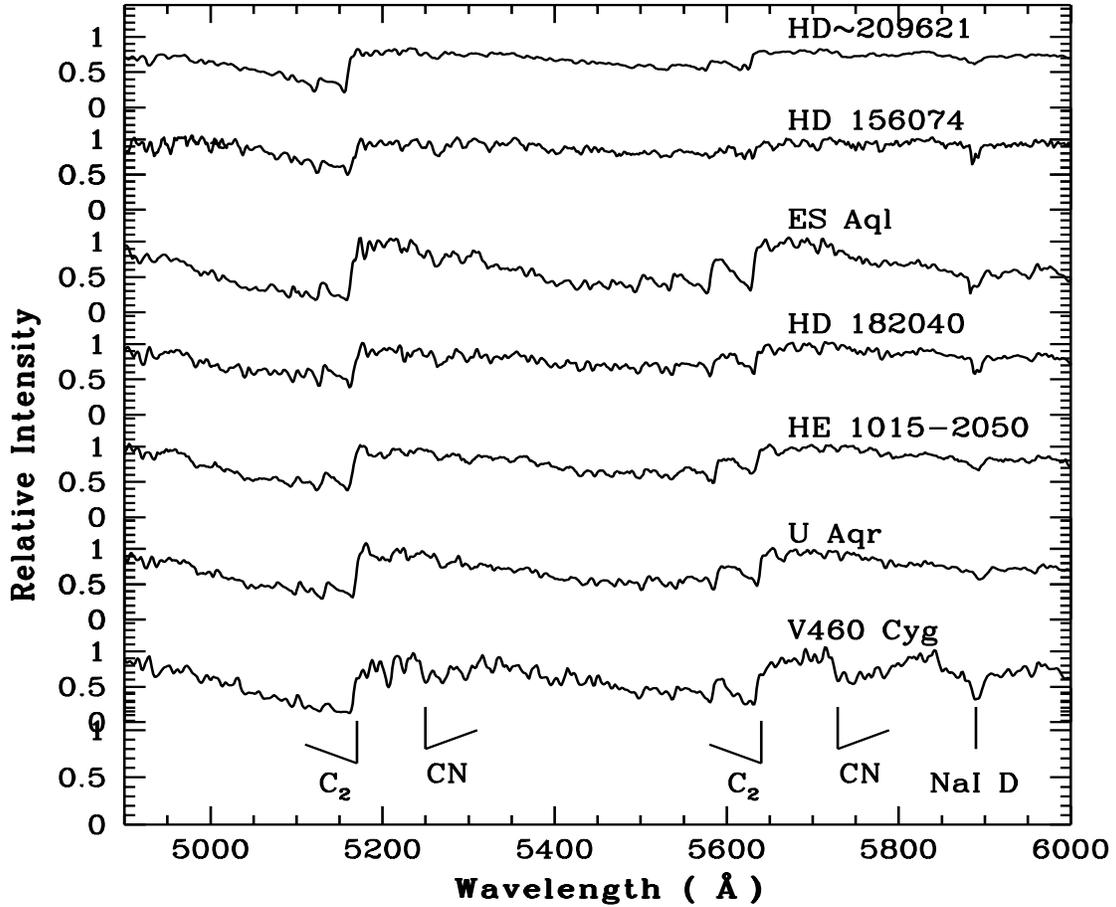}
\caption{ Same as Figure 1 except for the wavelength region 4900-6000 \AA\,. 
The weakness of CN bands relative to the C$_{2}$ bands in the spectra
of HE~1015$-$2050 and  other HdC and RCB stars is  noticeable. 
The features due to Na I D, although appearing  strong in
HE~1015$-$2050 \& U~Aqr, are much weaker relative to their counterparts
in V460~Cyg and HD~182040.  The HE star's spectrum bears a resemblance 
closest to  the spectrum of U~Aqr.
\label{fig2}
}
\end{figure*}

\begin{figure*}
\centering
\includegraphics[angle=0,height=13cm,width=15cm]{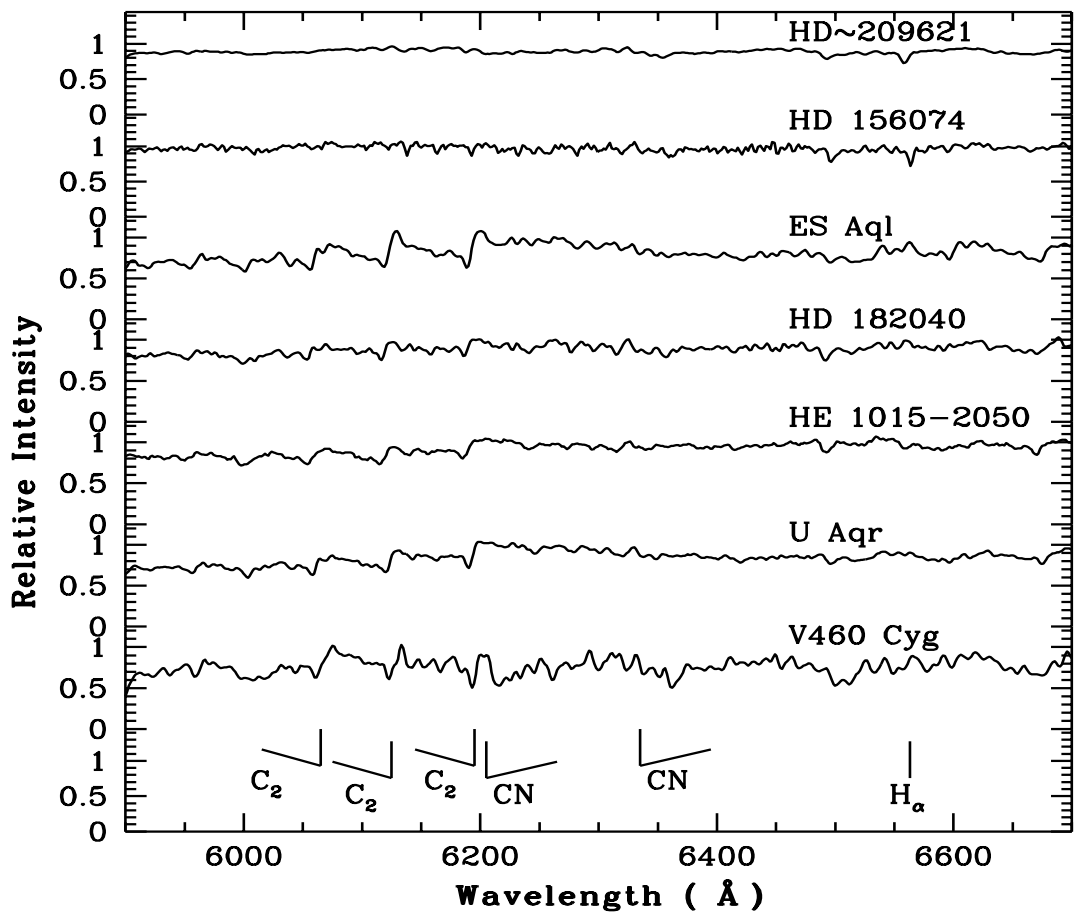}
\caption{Same as figure 1, except for the wavelength region 5900-6700 \AA\,.
The CN bands which appear with almost equal strengths in the  spectrum of
the CN star V460~Cyg are almost absent (or barely detectable) in the 
spectra of HE~1015$-$2050 and  U~Aqr. H$_{\alpha}$  
feature  is distinctly  seen in the spectra of the CH and C-R star 
HD~209621 and HD~156074 respectively. This feature is not detectable 
in the spectra of  HE~1015$-$2050,  HdC and RCB stars. Non detection of H$_{\alpha}$,
and marginal detection of G-band of CH  (figure 1) hint at hydrogen-poor 
nature of the objects.
\label{fig3}
}
\end{figure*}

\begin{figure*}
\centering
\includegraphics[angle=0,height=15cm,width=15cm]{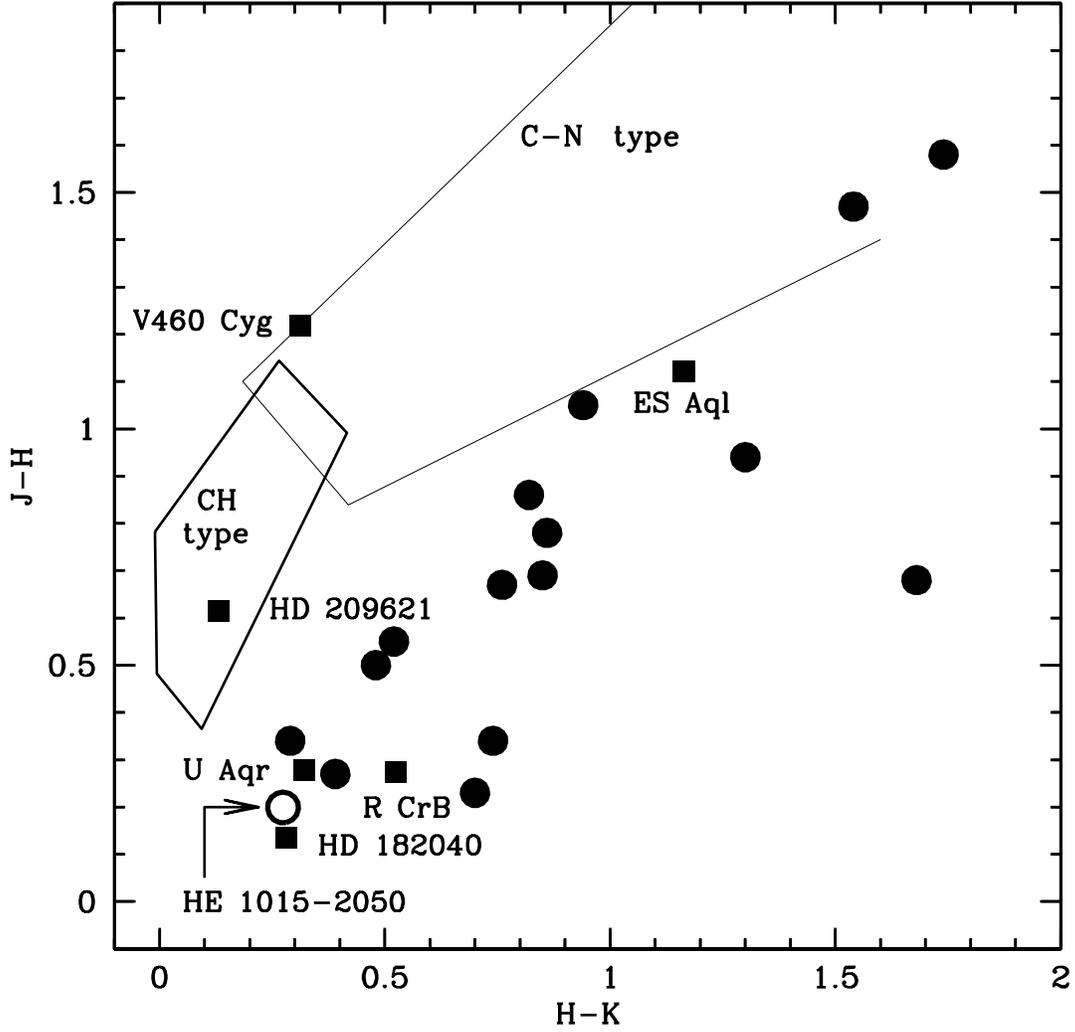}
\caption{
J-H versus H-K color magnitude diagram in which the  position of 
HE~1015$-$2050 is shown by an open circle. The colors of LMC RCB stars 
are shown with solid circles.  The positions of the comparison 
stars and the prototype RCB star R CrB are marked with solid squares. 
Most of the LMC RCB stars occupy 
a well defined region as noticed in the figure. The thick box on the 
lower left represents the location of CH stars and the thin box on the
upper right represents the location of C-N stars \citep{Totten00}. 
\label{fig4}
}
\end{figure*}

{\footnotesize
\begin{table*}
\centering
{\bf Table 1: Photometric parameters of HE~1015$-$2050 and the \\
comparison HdC and RCB stars  }\\
\tiny
\begin{tabular}{ccccccccccc}
\hline
             &              &                &          &            &       &      &            &         &       &      \\
Star No.     & RA(2000)     & DEC(2000)      & $l$      & $b$        & $B$   & $V$  & $B-V$      &  $J$    & $H$   & $K$    \\
             &              &                &          &            &       &      &            &         &       &       \\
\hline
             &              &                &          &            &       &      &            &         &       &       \\
HE~1015$-$2050 & 10 17 34.232 &$-$21 05 13.87  & 261.3144 &  29.0853   & 16.97 & 16.3 & 0.67$^{a}$ &  14.977 & 14.778& 14.504 \\
HE~2200$-$1652 & 22 03 19.690 &$-$16 37 35.29  & 39.1507  & $-$49.8124 & 12.16 & 11.17& 0.994      &  9.562  & 9.283 & 8.961  \\
( U~Aqr)     &              &                &          &            &       &      &            &         &       &        \\
HD~182040     & 19 23 10.077 & $-$10 42 11.54 & 26.8639  & $-$11.8517 &  8.02 &  7.0 &  1.07      &  5.364  & 5.228 & 4.947  \\
 ES~Aql      & 19 32 21.61  & $-$00 11 31.00 & 37.4640  & $-$09.1835 & 13.5  &  --- &  ---       &  10.244 & 9.123 & 7.960  \\
             &              &                &          &            &       &      &            &         &       &        \\
\hline
\end{tabular}

$^{a}$ From \citet{Christlieb01}\\

\end{table*}
}
\end{document}